\documentclass[conference,10pt]{IEEEtran}

\usepackage[letterpaper, left=0.61in, right=0.61in, top=0.65in, bottom=0.9in, includefoot]{geometry}

\pdfoutput=1
\usepackage[cmex10]{amsmath} 
\usepackage{cite,amsfonts,amssymb,psfrag,paralist}
\usepackage{cancel}
\usepackage{graphicx}
\usepackage{epstopdf}
\usepackage{rotating}
\usepackage{dsfont}
\usepackage{color}
\usepackage{tikz}
\usetikzlibrary{plotmarks}
\usepackage{pgfplots}
\usetikzlibrary{calc}
\usetikzlibrary{shapes,arrows}
\usetikzlibrary{decorations.markings}
\usetikzlibrary{positioning}
\pgfplotsset{compat=1.10}
\usetikzlibrary{calc}
\usetikzlibrary{shapes,arrows}
\usetikzlibrary{decorations.markings}

\usepackage{bbm}
\usepackage{balance}

\usepackage[utf8]{inputenc}
\usepackage[T1]{fontenc}
\usepackage{url}
\usepackage{ifthen}

\DeclareFontFamily{U}{mathx}{\hyphenchar\font45}
\DeclareFontShape{U}{mathx}{m}{n}{<-> mathx10}{}
\DeclareSymbolFont{mathx}{U}{mathx}{m}{n}
\DeclareMathAccent{\widebar}{0}{mathx}{"73}

\def\BibTeX{{\rm B\kern-.05em{\sc i\kern-.025em b}\kern-.08em T\kern-.1667em\lower.7ex\hbox{E}\kern-.125emX}}

\newcommand{\Enc}{\mathsf{Enc}}
\newcommand{\Dec}{\mathsf{Dec}}
\newcommand{\Est}{\mathsf{Est}}

\newcommand{\overbar}[1]{\mkern 1.5mu\overline{\mkern-1.5mu#1\mkern-1.5mu}\mkern 1.5mu}

\usepackage{scalerel}
\usepackage{stackengine}
\stackMath
\def\hatgap{2pt}
\def\subdown{-2pt}
\newcommand\reallywidehat[2][]{%
  \renewcommand\stackalignment{l}%
  \stackon[\hatgap]{#2}{%
    \stretchto{%
      \scalerel*[\widthof{$#2$}]{\kern-.6pt\bigwedge\kern-.6pt}%
      {\rule[-5\textheight]{0.1ex}{\textheight}}
    }{0.5ex}
    _{\smash{\belowbaseline[\subdown]{\scriptstyle#1}}}%
  }}

\newtheorem{theorem}{Theorem}

\newtheorem{definition}{Definition}

\begin{document}
\title{Low-latency Secure Integrated Sensing and Communication with Transmitter Actions}
\renewcommand\footnotemark{}
\renewcommand\footnoterule{}

\author{
  \IEEEauthorblockN{Truman Welling\textsuperscript{1}, Onur G\"unl\"u\textsuperscript{2}, and Aylin Yener\textsuperscript{1}}
    \IEEEauthorblockA{\textsuperscript{1}%
        Department of Electrical and Computer Engineering, The Ohio State University, USA
    }
    \IEEEauthorblockA{\textsuperscript{2}%
        Information Theory and Security Laboratory (ITSL), Link{\"o}ping University, Sweden\\ welling.78@osu.edu, onur.gunlu@liu.se, yener@ece.osu.edu
    }
}

\maketitle

\begin{abstract}
    This paper considers an information theoretic model of secure integrated sensing and communication, represented as a wiretap channel with action dependent states. This model allows securing part of a transmitted message against a sensed target that eavesdrops the communication, while enabling transmitter actions to change the channel statistics. An exact secrecy-distortion region is given for a physically-degraded channel. A finite-length achievability region is established for the model using an output statistics of random binning method, giving an achievable bound for low-latency applications.
\end{abstract}

\begin{IEEEkeywords}
integrated sensing and communication, wiretap channel, secure communications, finite blocklength analysis
\end{IEEEkeywords}

\section{Introduction}
Next generation connectivity will require both communication and sensing in applications such as self driving cars. As such, the integration of sensing and communication (ISAC) \cite{andersson2021joint} also allows for more efficient usage of bandwidth than independent sensing and communication \cite{NokiaGuysJCASTutorial}. This increase in efficiency comes at the expense of an inherent secrecy risk. If a target is sensed using an ISAC waveform, then the target has potentially observed the waveform, making the target also an eavesdropper.

Standard cryptographic methods are not a practical solution in many next generation systems, such as Internet of Things (IoT) settings where the devices have few computing resources or are designed for low power consumption. In these cases, physical layer security is a potential alternative \cite{yener2015pls}. Recently, information theoretic models have been proposed to explore the theoretical limits of physical layer security for ISAC channels \cite{MariMACJCAS,gunlu2023secureISAC,mittelbach2024secure,gunlu2024nonasymptotic,welling2024ISACAction}. In \cite{MariMACJCAS}, an ISAC channel was modeled as a state-dependent broadcast channel with feedback. Physical layer security was explored for this model in \cite{gunlu2023secureISAC}. A similar model characterizes the secrecy-distortion region for Rayleigh fading channels \cite{mittelbach2024secure}. All of these models assume that the channel output feedback is available to the transmitter while encoding.

The secure ISAC model in \cite{gunlu2024nonasymptotic} simplifies the model of \cite{gunlu2023secureISAC} by using the feedback for state estimation but not encoding. The lack of dependence in encoding on feedback allows the model to meet low-latency requirements. In this paper, we consider the setting of \cite{welling2024ISACAction} with the same simplification, not using feedback for action or channel encoding, allowing us to find results relevant in a low-latency setting. Our contributions are an exact secrecy-distortion region and an inner bound on the finite blocklength secrecy-distortion region.

First, the specific theoretical model is introduced following which we establish the exact asymptotic secrecy distortion region for the physically-degraded ISAC channel with transmitter actions. We then analyze the finite blocklength performance using nonasymptotic output statistics of random binning (OSRB) method \cite{AminOSRB,yassaeeNAOSRB} to derive an inner bound for the ISAC channel with transmitter actions.

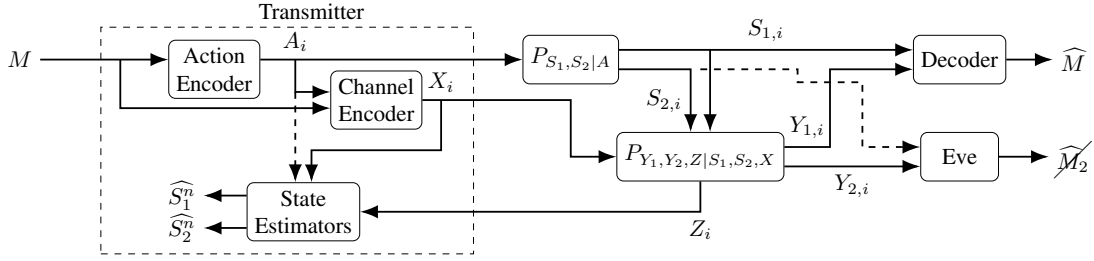
\begin{figure*}[ht]
  \centering
  \resizebox{0.8\linewidth}{!}{
    \begin{tikzpicture}
      \node (a) at (1.5,-.15) [draw,rounded corners = 3pt, minimum width=1.2cm,minimum height=0.75cm, align=center] {Action \\ Encoder};
      \node (b) at (7,0) [draw,rounded corners = 3pt, minimum width=1.2cm,minimum height=0.75cm, align=center] {$P_{S_{1},S_{2}|A}$};
      \node (c) at (4,-.625) [draw, rounded corners= 3pt, minimum width= 1.2cm, minimum height =0.75cm, align=center] {Channel \\ Encoder};
      \node (d) at (9,-1.5) [draw, rounded corners = 3pt, minimum width=1.2cm, minimum height=0.75cm, align=center] {$P_{Y_{1},Y_{2},Z|S_1,S_2,X}$};
      \node (e) at (13,0) [draw, rounded corners = 3pt, minimum width=1.2cm, minimum height=0.75cm, align=center] {Decoder};
      \node (f) at (13,-1.5) [draw, rounded corners = 3pt, minimum width=1.2cm, minimum height=0.75cm, align=center] {Eve};
      \node (g) at (2.875,-2.35) [draw, rounded corners = 3pt, minimum width=1.2cm, minimum height=0.75cm, align=center] {State\\ Estimators};
      \node (h) at (-1.5,0) {$M$};
      \node (i) at (14.75,0) {$\widehat{M}$};
      \node (j) at  (2.625,-1.25) [draw, dashed, minimum width=5.75cm, minimum height=3.5cm, align=center] {};
      \node (k) at (3,.75) {Transmitter};
      \draw[decoration={markings,mark=at position 1 with {\arrow[scale=1.5]{latex}}},
      postaction={decorate}, thick, shorten >=1.4pt] ($(h.east)+(0.0,0)$) -- ($(a.west)-(0,-0.15)$);
      \node (h1) at (0.05,0) {};
      \node (h2) at (0.05,-.75) {};
      \draw[decoration={markings,mark=at position 1 with {\arrow[scale=1.5]{latex}}},
      postaction={decorate}, thick, shorten >=1.4pt] ($(h1)$) -- ($(h2)$) -- ($(c.west)-(0,0.125)$);
      \draw[decoration={markings,mark=at position 1 with {\arrow[scale=1.5]{latex}}},
      postaction={decorate}, thick, shorten >=1.4pt] ($(a.east)+(0.0,0.15)$) -- ($(b.west)-(0,0)$);
      \node (a1) at (2.75,0) {};
      \node (a2) at (2.75,-.5) {};
      \draw[decoration={markings,mark=at position 1 with {\arrow[scale=1.5]{latex}}},
      postaction={decorate}, thick, shorten >=1.4pt] ($(a1) + (0,0)$) -- ($(a2)$) node [at start, above] {$A_i$} --  ($(c.west)+ (0,.125)$);
      \draw[decoration={markings,mark=at position 1 with {\arrow[scale=1.5]{latex}}},
      postaction={decorate}, thick, shorten >=1.4pt, dashed] ($(a2) + (0,0)$) -- ($(a2)$) --  ($(g.north)+ (-0.125,0)$);
      \node (c1) at (7,-.625) {};
      \node (c2) at (7,-1.5) {};
      \draw[decoration={markings,mark=at position 1 with {\arrow[scale=1.5]{latex}}},
      postaction={decorate}, thick, shorten >=1.4pt] ($(c.east) + (0,0)$)  -- ($(c1) + (0,0)$) -- ($(c2)$) -- ($(d.west)$);
      \node (c3) at (5,-.625) {};
      \node (c4) at (5,-1.4) {};
      \node (c5) at (3,-1.4) {};
      \draw[decoration={markings,mark=at position 1 with {\arrow[scale=1.5]{latex}}},
      postaction={decorate}, thick, shorten >=1.4pt] ($(c3) + (0,0)$) --  ($(c4)+ (0,0)$) node [at start, above] {$X_i$}  -- ($(c5)$) -- ($(g.north) + (.125,0)$);
      \node (b1) [above of = d, node distance = 1.5cm] {};
      \draw[decoration={markings,mark=at position 1 with {\arrow[scale=1.5]{latex}}},
      postaction={decorate}, thick, shorten >=1.4pt] ($(b.east) - (0,-.15)$) --  ($(e.west) - (0,-.15)$) node [midway, above] {$S_{1,i}$};
      \draw[decoration={markings,mark=at position 1 with {\arrow[scale=1.5]{latex}}},
      postaction={decorate}, thick, shorten >=1.4pt] ($(b1) - (-.15,-.15)$) --  ($(d.north) - (-.15,0)$);
      \draw[decoration={markings,mark=at position 1 with {\arrow[scale=1.5]{latex}}},
      postaction={decorate}, thick, shorten >=1.4pt] ($(b.east) + (0,-.15)$) -- ($(b1) + (-.15,-.15)$) --  ($(d.north) + (-.15,0)$) node [midway, left] {$S_{2,i}$};
      \node (b2) at (10.5,0) {};
      \node (b3) at (10.5,-.35) {};
      \node (f1) at (11.5,-.35) {};
      \node (f2) at (11.5,-1.5) {};
      \draw[decoration={markings,mark=at position 1 with {\arrow[scale=1.5]{latex}}},
      postaction={decorate}, thick, shorten >=1.4pt,dashed] ($(b1) + (0,-.15)$) -- ($(b2) + (0,-.15)$) -- ($(b3)$) -- ($(f1)$) --  ($(f2) + (0,.15)$) --  ($(f.west) + (0,.15)$);
      \node (d1) at (11,-1.5) {};
      \node (d2) at (11,0) {};
      \draw[decoration={markings,mark=at position 1 with {\arrow[scale=1.5]{latex}}},
      postaction={decorate}, thick, shorten >=1.4pt] ($(d.east) - (0,-.15)$) -- ($(d1) - (0,-.15)$) node [midway, above] {$Y_{1,i}$} -- ($(d2) + (0,-.15)$) --  ($(e.west) + (0,-.15)$);
      \draw[decoration={markings,mark=at position 1 with {\arrow[scale=1.5]{latex}}},
      postaction={decorate}, thick, shorten >=1.4pt] ($(d.east) + (0,-.15)$) --  ($(f.west) + (0,-.15)$) node [midway, below] {$Y_{2,i}$};
      \node (d3) at (9,-2.35) {};
      \draw[decoration={markings,mark=at position 1 with {\arrow[scale=1.5]{latex}}},
      postaction={decorate}, thick, shorten >=1.4pt] ($(d.south)$) --  ($(d3)$)  -- node [at start, below] {$Z_{i}$} ($(g.east) + (0,0)$);
      \draw[decoration={markings,mark=at position 1 with {\arrow[scale=1.5]{latex}}},
      postaction={decorate}, thick, shorten >=1.4pt] ($(e.east)$) --  ($(i.west)$);
      \node (g3) at (1,-2.1) {$\widehat{S_1^n}$};
      \node (g4) at (1,-2.6) {$\widehat{S_2^n}$};
      \draw[decoration={markings,mark=at position 1 with {\arrow[scale=1.5]{latex}}},
      postaction={decorate}, thick, shorten >=1.4pt] ($(g.west)+(0,.25)$) --  ($(g3.east)$);
      \draw[decoration={markings,mark=at position 1 with {\arrow[scale=1.5]{latex}}},
      postaction={decorate}, thick, shorten >=1.4pt] ($(g.west)-(0,.25)$) --  ($(g4.east)$);
      \node (f4) at (14.75,-1.5) {$\cancel{\widehat{M}_2}$};
      \draw[decoration={markings,mark=at position 1 with {\arrow[scale=1.5]{latex}}},
      postaction={decorate}, thick, shorten >=1.4pt] ($(f.east)$) -- ($(f4.west)$);
    \end{tikzpicture}
  }
  \vspace*{-0.2cm}
  \caption{Secure ISAC model with transmitter action-dependent states under partial secrecy, where we have $M=(M_1,M_2)$ and only $M_2$ should be kept secret from Eve, for $i~=~[1:n]$. The channel input $X_i$ is a function of $(M,A_i)$.}\label{fig:SecureStateDependentISACChannelModel}
  \vspace*{-0.3cm}
\end{figure*}

\section{Problem Definition}\label{sec:problem_definition}

Consider the channel model depicted in Figure~\ref{fig:SecureStateDependentISACChannelModel}, a state dependent broadcast channel with feedback, consisting of a transmitter who wants to communicate reliably with the legitimate receiver and sense a target while keeping part of the message secret from the target, thus also treating the target as an eavesdropper. The transmitter, observing a uniformly distributed message $M=(M_1,M_2)\in\mathcal{M}=\mathcal{M}_1\times\mathcal{M}_2$ computes the action sequence $A^n=\Enc_{Act}(M)\in\mathcal{A}^n$ and the channel input sequence $X^n=\Enc(M,A^n)\in\mathcal{X}^n$ where $\Enc_{Act}(\cdot)$ and $\Enc(\cdot,\cdot)$ are (random) encoding functions for the action and channel inputs, respectively. The states are determined by nature according to $(S_1^n,S_2^n)|\{A^n=a^n\}\sim\prod_{i=1}^nP_{S_1S_2|A}(s_{1,i},s_{2,i}|a_i)$. For each channel use $i\in[1:n]$, $(X_i,S_{1,i},S_{2,i})$ are inputs to the discrete memoryless channel $P_{Y_1Y_2|S_1S_2X}$ during which the legitimate receiver observes $Y_{1,i}\in\mathcal{Y}_1$ and $S_{1,i}\in\mathcal{S}_1$ and the target (eavesdropper) observes $Y_{2,i}\in\mathcal{Y}_2$ and $S_{2,i}\in\mathcal{S}_2$. After the $n^{th}$ transmission, the legitimate receiver forms an estimate $\widehat{M}$ of $M$ by $\widehat{M}=\Dec(Y_1^n,S_1^n)$ where $\Dec(\cdot,\cdot)$ is a decoding function. We denote the feedback to the transmitter by $Z_i\in\mathcal{Z}$. The sensing is abstracted as the estimates $(\widehat{S}_1^n,\widehat{S}_2^n)$ of the states, $\widehat{S}_j=\Est_j(A^n,X^n,Z^n)\in\widehat{\mathcal{S}}^n_j$ for $j=1,2$ where $\Est_j(\cdot,\cdot,\cdot)$ is an estimation function. All sets $\mathcal{A},\mathcal{X},\mathcal{Y}_1,\mathcal{Y}_2,\mathcal{S}_1,\mathcal{S}_2,\widehat{\mathcal{S}}_1,\widehat{\mathcal{S}}_2$ and $\mathcal{Z}$ are assumed to be finite.

In this model, the states represent the information that the sensor, which is at the transmitter, would like to approximate, such as angle of arrival. The transmitter actions abstract decisions at the transmitter that affect the distribution of the states, such as beamforming or physically moving. The feedback is the observation of the reflected communication waveform by the targets.
To simplify the analysis, we assume that the transmitter feedback is noiseless, i.e., $Z_i=(Y_{1,i},Y_{2,i})$ for all $i\in[1:n]$, which provides an outer bound for the performance of the noisy feedback scenario.

Now we define physical degradation for an ISAC channel with transmitter actions.
\begin{definition}\label{def:physicallydegraded}
  \normalfont An ISAC channel with transmitter actions as depicted in Fig.\ref{fig:SecureStateDependentISACChannelModel} is \emph{physically-degraded} if
  \begin{align}
    P_{AXY_1Y_2S_1S_2}&=P_{AX}P_{Y_1S_1Y_2S_2|AX}\nonumber\\
    &=P_{AX}P_{S_1|A}P_{Y_1|S_1X}P_{Y_2S_2|S_1Y_1}\label{eq:physicaldegradedcond}.
  \end{align}
  \hfill$\lozenge$
\end{definition}

\section{Asymptotic Performance Limits under Partial Secrecy} \label{sec:Asymptotic}

We first define an achievable tuple and the achievable secrecy distortion region.

\begin{definition}
    Under partial secrecy, a secrecy-distortion tuple of the form $(R_1,R_2,D_1,D_2)$, with $\log|\mathcal{M}_j|=nR_j$ for $j=1,2$ is achievable if for any $\delta>0$ there exists a channel encoder, action encoder, decoder, $n\geq1$, and two state estimators $\widehat{S}^n_j=\Est_j(X^n,A^n,Z^n)$ for $j=1,2$ such that
    \begin{align}
    &\Pr\big[(M_1,M_2) \neq (\widehat{M}_1,\widehat{M}_2)\big] \leq \delta&&\!\!\!\!\! (\text{reliability})\label{eq:asymptotic_reliability_constraint}\\
    &I(M_2;Y^n_2,S_2^n) \leq \delta&&\!\!\!\!\!(\text{strong secrecy})\label{eq:asymptotic_secrecyleakage_constaints}\\
    &\mathbb{E}\big[d_j(S_j^n,\widehat{S_j^n})\big] \!\leq\! D_j\!+\!\delta\;\;\;\;\;\;\text{for } j\!=\!1,2\;\;&&\!\!\!\!\!(\text{distortion})\label{eq:distortion_cons}
    \end{align}
    and we have per letter distortion metrics $d_j(\widehat{s}_j^n,s_j^n)=\frac{1}{n}\sum_{i=1}^nd_j(\widehat{s}_{j,i},s_{j,i})$ for $j=1,2$ that are bounded from above by some value $d_\text{max}$. The secrecy-distortion region $\mathcal{R}_{PS,Act}$ is the closure of all achievable tuples under partial secrecy.
    \hfill$\lozenge$ \label{def:asymptotic_achievable_tuple}
\end{definition}

Now we give the asymptotic secrecy-distortion region for a physically-degraded ISAC channel.
\begin{theorem}\label{theo:Asymptotic_SD_region}
  {\normalfont(Physically-degraded):} For a physically-degraded ISAC channel, $\mathcal{R}_{\textnormal{PS,Act}}$ is the union over all joint distributions $P_{VAX}$ of the rate tuples $(R_{1}, R_{2},D_1,D_2)$ satisfying
  \begin{align}
    R_{1}+R_2 &\leq I(V;Y_1,S_1)\label{eq:A_ach_rate1}\\
    R_{2} &\leq I(V;Y_1,S_1)-I(V;Y_2,S_2)\label{eq:A_ach_rate2}\\
    D_j &\geq \mathbb{E}[d_j(S_j,\widehat{S}_j))]\qquad\qquad  \text{for }j=1,2\label{eq:achdistortion1and2}
  \end{align}
  where we have
  \begin{align}
    &P_{VAXY_1Y_2S_1S_2}\!=\!P_{V|AX}P_{AX}P_{S_1|A}P_{Y_1|S_1X}P_{Y_2S_2|S_1Y_1}\label{eq:jointprobPD}
  \end{align}
  and one can use the deterministic per-letter estimators $\Est_j(a,x,y_1,y_2)=~\widehat{s}_j$ for $j=1,2$ such that
  \begin{align}
    &\Est_j(a,x,y_1,y_2)\nonumber\\
    &=\mathop{\textnormal{argmin}}_{\tilde{s}\in\widehat{\mathcal{S}}_j} \sum_{s_j\in\mathcal{S}_j}P_{S_j|AXY_1Y_2}(s_j|a,x,y_1,y_2)\; d_j(s_j,\tilde{s}).\label{eq:deterministicest}
  \end{align}
  One can also bound $|\mathcal{V}|$ by
  \begin{align}
    &|\mathcal{X}|\!\cdot\!|\mathcal{A}|+\!1.\label{eq:A_cardV}
  \end{align}\label{thm:asymptotic_rate_region}
\end{theorem}

The rate condition \eqref{eq:A_ach_rate1} signifies the combined rate of the common and secure messages should be below the reliable communication rate for the channel between the transmitter and legitimate receiver. \eqref{eq:A_ach_rate2} implies that the secure message rate is upper bounded by the wiretap coding rate, whereas the secure message rates in \cite{welling2024ISACAction} has an additional secret key rate.

\begin{IEEEproof}[Proof Sketch]
    The cardinality bound follows from \cite[Lemma 15.4]{CsiszarKornerbook2011}, where $|\mathcal{X}|\cdot|\mathcal{A}|-1$, letters are needed to preserve $P_{XA}$ with two additional letters to preserve $I(V;Y_1,S_1)$ and $I(V;Y_2,S_2)$.

    Achievability Proof: We use the OSRB method \cite{AminOSRB} for the achievability proof. To this end, we consider two problems, the channel coding problem assisted with shared randomness and the source coding problem that is operationally dual to the former. We find conditions for which the source coding protocol is reliable and secure, then we find conditions for which the probability distributions induced by the two protocols are arbitrarily close, providing a reliable, secure solution to the channel coding problem assisted with shared randomness. Finally, we remove the shared randomness by showing that the protocol remains secure and reliable for a specific realization of the shared randomness.

    Fix $P_{VAX}(v,a,x)$ for which there exist per-letter estimators $\Est_j(A,X,Y_1,Y_2)=\widehat{S}_j$ for $j=1,2$ that satisfy $\mathbb{E}\big[d_j(S_j,\widehat{S}_j)\big]\leq D_j+\epsilon_n$ for $j=1,2$ and $0<\epsilon_n\rightarrow 0$ as $n\rightarrow\infty$.

    \emph{Protocol A:} We first generate a tuple of random variables $(V^n,A^n,X^n,Y_1^n,S_1^n,Y_2^n,S_2^n)$ according to $P_{VAXY_1S_1Y_2S_2}$ as defined in \eqref{eq:jointprobPD}. The source encoder, observing $V^n$, uniformly and independently assigns random bin indices $\mathcal{B}_{M_1}(V^n)=M_1\in\mathcal{M}_1=[1:2^{nR_1}]$, $\mathcal{B}_{M_2}(V^n)=M_2\in\mathcal{M}_1=[1:2^{nR_2}]$, and $\mathcal{B}_{F}(V^n)=F\in[1:2^{n\widetilde{R}}]$.

    The random binning induces the following random pmf
    \begin{align}
        P^{RB}\!=\!P^{RB}_{M_1M_2F}P^{RB}_{V^n|M_1M_2F}P_{A^nX^nY_1^nS_1^nY_2^nS_2^n|V^n}.\label{eq:RB_pmf}
    \end{align}
    \emph{Protocol B:} We assume $F$ is uniformly and independently selected from $[1:2^{n\widetilde{R}_v}]$ and is shared with all parties prior to transmission. The transmitter selects a message $M=(M_1,M_2)$ uniformly from $(\mathcal{M}_1\times\mathcal{M}_2)$ independent of  $F$. Using $P^{RB}_{V^n|M_1M_2F}$ from Protocol A, the transmitter generates $V^n$. The channel inputs $(A_i,X_i)$ for the $i^{th}$ channel use are produced by the transmitter according to $P_{XA|V}$.

    The pmf induced by Protocol B is
    \begin{align}
        P^{RC}\!\!=\!P^U_{M_1M_2}P^U_FP^{RB}_{V^n|M_1M_2F}P_{A^nX^nY_1^nS_1^nY_2^nS_2^n|V^n}.\label{eq:RC_pmf}
    \end{align}

    A sufficient condition reliable decoding using a Slepian-Wolf decoder in Protocol A \cite[Lemma 1]{AminOSRB} is
    \begin{align}
        \widetilde{R}>H(V|Y_1,S_1). \label{eq:asymp_reliability_suffcond}
    \end{align}
    Privacy amplification \cite[Theorem 1]{AminOSRB} ensures almost independence of $M_2$ and $F$ with the eavesdropper's observation $(Y_1^n,S_1^n)$ for Protocol A if
    \begin{align}
        \widetilde{R}_v+R_2 < H(V|Y_2,S_2) \label{eq:asymp_secrecy_suffcond}
    \end{align}
    establishing security for Protocol A. Finally, using \cite[Theorem1]{AminOSRB}, the induced distributions \eqref{eq:RB_pmf} and \eqref{eq:RC_pmf} are arbitrarily close in variational distance when
    \begin{align}
        R_1+R_2+\widetilde{R}_v <H(V) \label{eq:asymp_approximate_suffcond}
    \end{align}
    which gives the desired reliability and security guarantees for Protocol B. A coding scheme for the original channel coding problem is obtained by finding a specific realization f of the common randomness $F$ for which the reliability and security conditions continue to hold as in \cite{AminOSRB}.

    Using Fourier-Motzkin elimination \cite{FMEbook} to simplify \eqref{eq:asymp_reliability_suffcond}-\eqref{eq:asymp_approximate_suffcond} gives \eqref{eq:A_ach_rate1} and
    \begin{align}
        R_2 &< \big[I(V;Y_2,S_2)-I(V;Y_1,S_1)\big]^+ \label{eq:asymp_r2_intermediate}
    \end{align}
    where $[a]^+=\max(0,a)$.
    Application of the Markov chain $V-(A,X)-(Y_1,S_1)-(Y_2,S_2)$, which follows from channel degradation, and the data processing inequality to \eqref{eq:asymp_r2_intermediate} recovers \eqref{eq:A_ach_rate2}. The distortion constraints in \eqref{eq:achdistortion1and2} and the viability of the deterministic per-letter estimators in \eqref{eq:deterministicest} follow as in \cite{gunlu2023secureISAC,welling2024ISACAction}.

    \vspace{.5em}
    Converse Proof: Suppose for some $n\geq1$, $\delta_n>0$, there exists a channel encoder, action encoder, decoder, and state estimators such that
    \begin{align}
        P\{M\neq\widehat{M}&\leq\delta_n \label{eq:asymp_converse_reliability}\\
        I(M_2;Y_2^n,S_2^n) &\leq\delta_n \label{eq:asymp_converse_secrecy}\\
        \mathbb{E}\big[d_j(S_j^n,\widehat{S}_j^n\big] &\leq D_j + \delta_n \label{eq:asymp_converse_distortion}
    \end{align}are satisfied for some tuple $(R_1,R_2,D_1,D_2)$.

    We have
    \begin{align}
        H(M_1,M_2|Y_1^n,S_1^n)\overset{(a)}{\leq} H(M_1,M_2|\widehat{M}_1,\widehat{M}_2) \overset{(b)}{\leq} n\epsilon_n \label{eq:asymp_converse_fano}
    \end{align}
    where $(a)$ follows by allowing randomized decoding and $(b)$ follows from Fano's inequality with $\epsilon_n=\frac{H_b(\delta_n)}{n}+\delta_n(R_1+R_2)$. Note that $\epsilon_n\rightarrow0$ as $\delta_n\rightarrow0$.

    We also define $V_i\triangleq(M_1,M_2,Y_1^{i-1},S_1^{i-1},Y_2^{i-1},S_2^{i-1})$ such that the Markov chain $V_i-(A_i,X_i)-(Y_{1,i},S_{1,i},Y_{2,i},S_{2,i})$ holds.

    \noindent\emph{Bound on $n(R_1+R_2)$}:
    \begin{align}
        &n(R_1+R_2) \overset{(a)}{\leq}I(M_1,M_2;Y_1^n,S_1^n) + n\epsilon_n\nonumber\\
        &\leq\sum\limits_{i=1}^n\big[H(Y_{1,i}S_{1,i})\nonumber\\
        &\quad-H(Y_{1,i},S_{1,i}|M_1,M_2,Y_1^{i-1},S_1^{i-1},Y_2^{i-1},S_2^{i-1})\big]+n\epsilon_n\nonumber\\
        &\overset{(b)}{=} \sum\limits_{i=1}^nI(V_i;Y_{1,i},S_{1,i}) + n\epsilon_n \label{eq:asymp_converse_r1pr2}
    \end{align}
    where $(a)$ follows from \eqref{eq:asymp_converse_fano} and $(b)$ follows from the definition of $V_i$.

    \noindent\emph{Bound on $nR_2$}:
    \begin{align}
        &nR_2 \overset{(a)}{\leq} H(M_2|Y_2^n,S_2^n) + \delta_n\nonumber\\
        & = I(M_2;Y_1^n,S_1^n|Y_2^n,S_2^n)+H(M_2|Y_1^n,Y_2^n,S_1^n,S_2^n) + \delta_n\nonumber\\
        & \leq I(M_2;Y_1^n,S_1^n|Y_2^n,S_2^n)+H(M_1,M_2|Y_1^n,S_1^n) + \delta_n\nonumber\\
        & \overset{(b)}{\leq} I(M_2;Y_1^n,S_1^n|Y_2^n,S_2^n)+n\epsilon_n + \delta_n\nonumber\\
        & \overset{(c)}{\leq}\sum\limits_{i=1}^n\big[H(Y_{1,i},S_{1,i}|Y_{2,i},S_{2,i})\nonumber\\
            & \quad-H(Y_{1,i},S_{1,i}|Y_1^{i-1},S_1^{i-1},M_1,M_2,Y_2^i,S_2^i)\big] + n\epsilon_n+\delta_n\nonumber\\
        & \overset{(d)}{=}\sum\limits_{i=1}^n\big[I(V_i;Y_{1,i},S_{1,i}|Y_{2,i},S_{2,i})\big] + n\epsilon_n +\delta_n\nonumber\\
        & \overset{(e)}{=}\sum\limits_{i=1}^n\big[I(V_i;Y_{1,i},S_{1,i})-I(V_i;Y_{2,i},S_{2,i})\big]\!+\!n\epsilon_n\!+\!\delta_n \label{eq:asymp_converse_r2}
    \end{align}
    where $(a)$ follows from \eqref{eq:asymp_converse_secrecy}, $(b)$ follows from \eqref{eq:asymp_converse_fano}, $(c)$ follows by the chain rule and application of the Markov chain $(Y_{1,i},S_{1,i})-(Y_1^{i-1},S_1^{i-1},M_1,M_2,Y_2^i,S_2^i)-(Y_{2,i+1}^n,S_{2,i+1}^n)$, $(d)$ follows from the definition of $V_i$, and $(e)$ follows because of the channel degradation \eqref{eq:physicaldegradedcond}.

    By introducing a time sharing random variable Q uniformly distributed over $[1:n]$ and independent of all other random variables and following the standard time sharing argument applied to \eqref{eq:asymp_converse_r1pr2} and \eqref{eq:asymp_converse_r2}, letting $V=(V_Q,Q)$, $X=X_Q$, $A=A_Q$, $Y_1=Y_{1,Q}$, $S_1=S_{1,Q}$, $Y_2=Y_{2,Q}$, and $S_2=S_{2,Q}$ such that $V-(A,X)-(Y_1,S_1)-(Y_2,S_2)$ forms a Markov chain, we recover the rate conditions \eqref{eq:A_ach_rate1} and \eqref{eq:A_ach_rate2}.
\end{IEEEproof}

\section{Inner Bound on Nonasymptotic Performance under Partial Secrecy} \label{sec:Nonasymptotic}

We now consider the achievable performance of our model for a finite block length $n$. We first adjust Definition \ref{def:asymptotic_achievable_tuple} to a fixed blocklength $n$. Following that we define some quantities that will be useful in the statement and proof of the nonasymptotic achievable region.

\begin{definition}
    Under partial secrecy and for fixed $\delta_r,\delta_D,\delta_\text{sec}>0$ and $n\geq1$, a nonasymptotic secrecy-distortion tuple $(R_1,R_2,D_1,D_2)$, with $\log|M_j|=nR_j$ for $j=1,2$, is $(\delta_r,\delta_D,\delta_\text{sec},n)$-achievable if there exists an action encoder, channel encoder, decoder, and two per-letter state estimators $\widehat{S}_j=\Est_j(X,A,Y_1,Y_2)$ such that we have \eqref{eq:distortion_cons} with per-letter distortion metrics bounded by a value $d_\text{max}$ and
    \begin{align}
    &\Pr\big[(M_1,M_2) \neq (\widehat{M}_1,\widehat{M}_2)\big] \leq \delta_r&&\!\!\!\!\! (\text{reliability})\label{eq:nonasymptotic_reliability_constraint}\\
    &\parallel P_{M_2Y_2^nS_2^n}-P^U_{M_2}P_{S_2Y_2}^n \parallel_1 \leq \delta_\text{sec}&&\!\!\!\!\!(\text{strong secrecy})\label{eq:nonasymptotic_secrecyleakage_constaints}\\
    &\;\mathbb{E}\big[d_j(S_j^n,\widehat{S_j^n})\big] \!\leq\! \delta_D\;\;\;\;\;\;\text{for } j\!=\!1,2\;\;&&\!\!\!\!\!(\text{distortion})\label{eq:nonasymp_distortion_constraints}.
    \end{align}
    The nonasymptotic secrecy-distortion region $\mathcal{R}_{Act}(\delta_r,\delta_d,\delta_\text{sec},n)$ is the closure of all $(\delta_r,\delta_d,\delta_\text{sec},n)$-achievable tuples under partial secrecy.
    \hfill$\lozenge$
\end{definition}

We define the information of a probability distribution $P_A$ as
\begin{align}
    h_{P_A} = \log\frac{1}{P_A(a)}
\end{align}
and the information density of a distribution $P_{AB}$ as
\begin{align}
    \imath(A,B)=\log\frac{P_{AB}(a,b)}{P_A(a)P_B(b)}.
\end{align}
Next, we define the dispersion of the channels
\begin{align}
    \mathcal{V}_{Y_1S_1} =\min_{P_{V|Y_1S_1}}\big[\text{Var}_{P_{VY_1S_1}}[\imath(V,Y_1S_1)|V]\big]\\
    \mathcal{V}_{Y_2S_2} =\min_{P_{V|Y_2S_2}}\big[\text{Var}_{P_{VY_2S_2}}\big[\imath(V,Y_2S_2)|V]\big]
\end{align}
where $\text{Var}[\cdot]$ is the variance. $Q(\cdot)$ denotes the standard normal tail probability. Define
\begin{align}
    \mu_{s\widehat{s}}=\min_{(s\widehat{s})\in\text{supp}(P_{S\widehat{S}})}P_{S\widehat{S}}(s,\widehat s).
\end{align}

We next establish an inner bound on the nonasymptotic secrecy-distortion region for the secure ISAC model considered.
\begin{theorem}
    For an ISAC channel with transmitter actions, a $(R_1,R_2,D_1,D_2)$ tuple is $(\delta_r,\delta_D,\delta_\text{sec},n)$-achievable if, for any $\theta\in[0,1]$, we have
    \begin{align}
        &R_1+R_2 \leq \bigg[I(V;Y_1,S_1) -\mathcal{O}\bigg(\frac{\log n}{n}\bigg)\nonumber\\
        &\qquad\qquad -Q^{-1}\bigg(\theta\Big(\delta_r+\mathcal{O}\Big(\frac{1}{\sqrt{n}}\Big)\Big)\bigg)\sqrt{\frac{\mathcal{V}_{Y_1S_1}}{n}}\bigg]^+ \label{eq:nonasymp_ach_r1pr2}\\
        &R_2 \leq \bigg[I(V;Y_1,S_1)-I(V;Y_2,S_2) -\mathcal{O}\bigg(\frac{\log n}{n}\bigg)\nonumber\\
        &\qquad\qquad -Q^{-1}\bigg((1-\theta)\Big(\delta_\text{sec}+\mathcal{O}\Big(\frac{1}{\sqrt{n}}\Big)\Big)\bigg)\sqrt{\frac{\mathcal{V}_{Y_2S_2}}{n}}\nonumber\\
        &\qquad\qquad -Q^{-1}\bigg(\theta\Big(\delta_r+\mathcal{O}\Big(\frac{1}{\sqrt{n}}\Big)\Big)\bigg)\sqrt{\frac{\mathcal{V}_{Y_1S_1}}{n}}\bigg]^+\label{eq:nonasymp_ach_r2}\\
        &D_j \geq \mathbb{E}[d_j(S_j,\widehat{S}_j)]-\epsilon_D  \qquad \text{for }j=1,2
    \end{align}
    such that
    \begin{align}
        \delta_D=\epsilon_D(1+D_j+\epsilon_D)+2|\mathcal{S}||\widehat{\mathcal{S}}|e^{-2n\epsilon_D^2\mu_{s_j\widehat{s}_j}}d_\text{max}
    \end{align}
    where we have
    \begin{align}
        &P_{VAXY_1Y_2S_1S_2}\!=\!P_{V|AX}P_{AX}P_{S_1S_1|A}P_{Y_1Y_2|S_1S_2X}\label{eq:nonasymp_joint_pmf}
    \end{align}
    and the per-letter estimators in \eqref{eq:deterministicest}.\label{thm:nonasymptotic_achievability}
\end{theorem}
\begin{IEEEproof}[Proof Sketch]
    The proof sketch will proceed as follows:
    \begin{enumerate}
        \item We define Protocols A and B and we establish a bound on the reliability and security.
        \item We combine the bounds on reliability and security to bound the expected variational distance between the distribution induced by the channel coding problem and the distribution that satisfies the reliability and security constraints.
        \item We establish rate conditions such that the reliability and security constraints are satisfied for $\theta\delta_r$ and $(1-\theta)\delta_\text{sec}$, for any $\theta\in[0,1]$. The rate conditions come from the minimization of the measure of atypical sets using the Berry-Esseen CLT.
    \end{enumerate}

    Fix a distribution $P_{VAX}$ such that $\mathbb{E}[d_j(S_i,\widehat{S}_j)]\leq D_j\!+\!\epsilon_D$.

    \emph{Protocol A:} We define the random binning as in the proof of Theorem~\ref{thm:asymptotic_rate_region} with the exception that we use \eqref{eq:nonasymp_joint_pmf} and the stochastic likelihood decoder, as defined in \cite{yassaeeNAOSRB,cerviaFixedStrongCoordination}, associated with the probability distribution
    \begin{align}
        T(\widehat{v}^n|y_1^n,s_1^n,f)=\frac{t(\widehat{v}^n|y_1^n,s_1^n)\mathds{1}\{\mathcal{B}_F(\widehat{v}^n)=f\}}{\sum\limits_{\overbar{v}^n\in\mathcal{V}^n}t(\overbar{v}^n|y_1^n,s_1^n)\mathds{1}\{\mathcal{B}_F(\overbar{v}^n)=f\}}
    \end{align}
    using an arbitrary pmf $t_{\widehat{V}^nY_1^nS_1^n}$ selected by the legitimate receiver.

    \noindent\emph{Protocol B:} We define Protocol B as in the proof of Theorem~\ref{thm:asymptotic_rate_region} with the exception of the decoder, for which we choose the stochastic likelihood decoder associated with $T_{\widehat{V}^n|Y_1^nS_1^nF}$ used in Protocol A.

    We define the following atypical sets. For $\gamma_1,\gamma_2,\gamma_3$ positive real numbers, we have
    \begin{align}
        &S_{\gamma_1}\!=\!\{v^n: h_{P_{V^n}}(v^n)\!-\!n(R_1+R_2+\widetilde{R})\!>\!\gamma_1\}\label{eq:s1_def}\\
        &S_{\gamma_2}\!=\!\{(v^n,y_1^n,s_1^n): n\widetilde{R}\!-\!h_t(v^n|y_1^n,s_1^n)\!>\!\gamma_2\}\label{eq:s2_def}\\
        &S_{\gamma_3}\!=\!\{(v^n,y_2^n,s_2^n): h_{P_V^n|Y_2^nS_2^n}(v^n|y_2^n,s_2^n)\!-\!n(R_2\!+\!\widetilde{R})\!>\!\gamma_3\}.\label{eq:s3_def}
    \end{align}

    We first establish a bound on the probability of error of Protocol A. The expected variational distance between the distributions induced by Protocols A and B is bounded by \cite[Theorem 1]{yassaeeNAOSRB}, i.e.,
    \begin{align}
        \mathbb{E}\big\|P^{RB}-P^{RC}\big\|_1\leq \epsilon_\text{apx}
    \end{align}
    where $\epsilon_\text{apx}=P_{V^n}(S_{\gamma_1^c})+2^{-\frac{1}{2}(\gamma_1+1)}$.
    By \cite[Theorem 2]{yassaeeNAOSRB}, the probability of error for the Protocol A is bounded by $\epsilon_\text{dec}$ where where $\epsilon_\text{dec}=P_{V^nY_1^nS_1^n}(S_{\gamma_2^c})+2^{-\gamma_2}$. Then for Protocol B, we have
    \begin{align}
        \mathbb{E}\big[P^{RC}\{V^n\neq\widehat{V}^n\}\big]< \epsilon_\text{apx} + \epsilon_\text{dec}.\label{eq:nonasymp_ach_proberror_edec}
    \end{align}
    The triangle inequality and \cite[Theorems 1 and 2]{cerviaFixedStrongCoordination} give
    \begin{align}
        &\mathbb{E}\big\| \!P^{RC}_{M_2FY_2^nS_2^n}\!-\!P_{M_2}^U\!P_F^UP_{Y_2^nS_2^n}\!\big\|_1\! \leq \epsilon_\text{apx} + \epsilon_\text{sec} \label{eq:nonasymp_ach_sec_initial_cond}
    \end{align}
    where $\epsilon_\text{sec}=P_{V^nY_2^nS_2^n}(S_{\gamma_3^c})+2^{-\frac{1}{2}(\gamma_3+1)}$. We let $\theta\delta_r=\epsilon_\text{apx}+\epsilon_\text{dec}$ and $(1-\theta)\delta_\text{sec}= \epsilon_\text{apx} + \epsilon_\text{sec}$.

    Now we combine the bounds on reliability and security. By the triangle inequality and $(b)$ follows from \eqref{eq:nonasymp_ach_proberror_edec} and \cite[Proposition 1]{cerviaFixedStrongCoordination} and \eqref{eq:nonasymp_ach_sec_initial_cond} we have
    \begin{align}
        &\mathbb{E}\big\| P^{RC}_{M_2FY_2^nS_2^n\widehat{V}^n}-P_{M_2}^UP_F^UP_{Y_2^nS_2^n}\mathds{1}\{\widehat{V}^n=V^n\}\big\|_1 \nonumber\\
        &\leq 2\epsilon_\text{apx} + \epsilon_\text{sec} + 4\epsilon_\text{dec} = \epsilon_\text{tot}.\label{eq:nonasymp_tot_bound}
    \end{align}
    We now eliminate the common randomness $F$ by conditioning on a specific realization $f$. Specifically, \cite[Lemma 3]{AminOSRB} allows us to bound \eqref{eq:nonasymp_tot_bound} by $2\epsilon_\text{tot}$ given $F=f^*$ for some $f^*$.

    We now establish the rate conditions through minimizing $2\epsilon_\text{tot}$ via the following inequality
    \begin{align}
        2\epsilon_\text{tot}& \leq 32(P_{V^nY_1^nS_1^nY_2^nS_2^n}\big((S_{\gamma_1}\cap S_{\gamma_2}\cap S_{\gamma_3})^c\big)\nonumber\\
        &\qquad+ 2\big(2^{-\frac{1}{2}(\gamma_1+1)}+4\cdot2^{-\gamma_2}+2^{-\frac{1}{2}(\gamma_3+1)}\big)
    \end{align}
    which follows by noting $P_{V^nY_1^nS_1^nY_2^nS_2^n}(S_{\gamma_1}^c\cup S_{\gamma_2}^c\cup S_{\gamma_3}^c)>P_{V^nY_1^nS_1^nY_2^nS_2^n}(S_{\gamma_j}^c\cap S_{\gamma_i}^c)$ for $i\neq j$ and an application of De Morgan's laws. What remains is to minimize the measure of the set $(S_{\gamma_1}\cap S_{\gamma_2}\cap S_{\gamma_3})^c$ and prudent selection of $\gamma_1$, $\gamma_2$, and $\gamma_3$.

    Following the steps in \cite{cerviaFixedStrongCoordination} and letting $t(v^ny_1^ns_1^n)=\prod_{i=1}^nP_{V|Y_1S_1}(v_i,y_{1,i},s_{1,i})$, we have that bounding the measure of $(S_{\gamma_1}\cap S_{\gamma_2}\cap S_{\gamma_3})^c$ is equivalent to bounding the measure of $S_1\cup S_2$ where
    \begin{align}
        &S_1\!=\!\{(v^n\!,y_1^n\!,s_1^n\!,y_2^n\!,s_2^n)\!: \!n(R_1+R_2)\!\geq\! \imath(v^n,y_1^ns_1^n)\!-\!\gamma_1\!-\!\gamma_2\}\nonumber\\
        &S_2\!=\!\{(v^n\!,y_1^n\!,s_1^n\!,y_2^n\!,s_2^n)\!:\nonumber\\
        &\qquad\qquad \!nR_2\!\geq\! \imath(v^n\!,y_1^ns_1^n\!)\!-\!\imath(v^n\!,y_2^ns_2^n\!)\!-\!\gamma_2\!-\!\gamma_3\}.
    \end{align}

    We outline the bound on $P(S_1\cup S_2)$. Since the tuple of random variables $(v^n,y_1^n,s_1^n,y_2^n,s_2^n)$ are i.i.d., the terms $Z_i=\imath(v_i,y_{1,i}s_{1,i})$ are also independent for $i=1,\dots,n$. Defining $\mu_n=1/n\sum_{i=1}^n\mathbb{E}[Z_i]$ and $V_n=1/n\sum_{i=1}^n\text{Var}[Z_i]$, if we assume that
    \begin{align}
        n(R_1+R_2)<n\mu_n-Q^{-1}(\theta\delta_r)n\sqrt{\frac{V_n}{n}}-\gamma_1-\gamma_2 \label{eq:nonasymp_inital_condition}
    \end{align}
    then we can define the set
    \begin{align}
        \bigg\{\!(v^n,y_1^n,s_1^n,y_2^n,s_2^n):\!\sum\limits_{i=1}^n\!\imath(V;Y_1S_1)\!<\!n\Big(\!\mu_n\!\! -\! t\sqrt{\frac{V_n}{n}}\Big)\!\bigg\} \label{eq:s1_outerbound}
    \end{align}
    which contains $S_1$. Thus bounding \eqref{eq:s1_outerbound} will also bound $S_1$. Application of the Berry Esseen CLT, see \cite{yassaeeNAOSRB}, gives that the measure of \eqref{eq:s1_outerbound} is upper bounded by $\theta\delta_r+\mathcal{O}\big(\frac{1}{\sqrt{n}}\big)$, we obtain $P(S_1)\leq\theta\delta_r+\mathcal{O}\big(\frac{1}{\sqrt{n}}\big)$. Since $(v_i,y_{1,i},s_{1,i})$ are generated i.i.d. according to $P_{VY_1S_1}$ for all $i=1,\dots,n$, we have
    \begin{align}
        \mu_n & =I(V;Y_1,S_1)\\
        V_n &\geq \mathcal{V}_{V|Y_1S_1}
    \end{align}
    which in combination with prudent selection of $(\gamma_1,\gamma_2)$ results in \eqref{eq:nonasymp_ach_r1pr2}. The bound on $P(S_2)$ follows similarly, resulting in \eqref{eq:nonasymp_ach_r2}. The distortion constraints follow as in the proof of \cite[Theorem~1]{gunlu2024nonasymptotic}.
\end{IEEEproof}

\section{Conclusion}

Theorem \ref{thm:asymptotic_rate_region} provides the exact asymptotic result for the secure ISAC model considered, whereas Theorem \ref{thm:nonasymptotic_achievability} provides an achievable region for low-latency secure ISAC with a fixed blocklength and specific bounds on the reliability, strong secrecy, and distortion measures. While the results of Theorem \ref{thm:asymptotic_rate_region} assume that the channel is physically degraded, Theorem \ref{thm:nonasymptotic_achievability} is more general and holds for all channels described by Fig. \ref{fig:SecureStateDependentISACChannelModel}.

Our work here can be extended through establishing a finite blocklength converse corresponding to Theorem~\ref{thm:nonasymptotic_achievability}. Other possible extensions include considering more sophisticated eavesdroppers or how these results scale in more complicated networks, such as one with multiple eavesdroppers.

\section*{Acknowledgment}
This work was supported in part by the by the U.S. Department of Transportation under Grant 69A3552348327 for the CARMEN+ University Transportation Center, ZENITH Research and Leadership Career Development Fund, Chalmers Transport Area of Advance, and the ELLIIT funding.

\bibliographystyle{IEEEtran}
\bibliography{references}

\end{document}